\documentclass[journal]{new-aiaa}

\usepackage[utf8]{inputenc}

\usepackage{graphicx}
\usepackage[version=4]{mhchem}
\usepackage{siunitx}
\usepackage{natbib}
\usepackage{longtable,tabularx}
\usepackage{xfrac}
\usepackage{multicol}
\usepackage{blindtext}
\usepackage{multirow} 
\usepackage{tikzpagenodes}
\usepackage{color} 
\usepackage{cancel} 

\newlength{\figwidth}
\newcommand\Rey{\mbox{\textit{Re}}} 
\newcommand\Nu{\mbox{\textit{Nu}}} 
\newcommand\Pra{\mbox{\textit{Pr}}}

\title{Minimum-Mass Limits for Streamlined Venus Atmospheric Probes}

\author{Jacob S. Izraelevitz \footnote{Robotics Technologist, Mobility and Robotic Systems Section. jacob.izraelevitz@jpl.nasa.gov} and Jeffery L. Hall \footnote  {Assistant Section Manager, Mobility and Robotic Systems Section. jeffery.l.hall@jpl.nasa.gov. Associate Fellow, AIAA}}
\affil{NASA Jet Propulsion Laboratory, California Institute of Technology, Pasadena CA 91109}

\begin{document}
	

\maketitle

\begin{abstract}
	Small, expendable drop probes are an attractive method for making measurements in the lower atmosphere of Venus, \,augmenting the capabilities of orbiters or aerial platforms that must remain in the benign temperature region above 50 km altitude. However, probe miniaturization is impeded by the need to provide thermal and pressure protection for conventional payloads. This paper determines the minimum mass limits for an insulated pressure vessel probe that operates all the way to the Venusian surface. Scaling laws for the probe performance and mass of major system components are explicitly derived using a simple model that captures the relevant physics. Streamlining the probe is found to be a highly effective strategy for lowering the system mass, but it also reduces the time available for data collection and transmission. Tradeoffs, guidelines and design charts are presented for an array of miniaturized probes. Total system masses on the order of 5 kilograms are plausible with streamlined probes if the desired science measurements can be performed faster than a standard Venus descent timeline.
\end{abstract} 

 \begin{tikzpicture}[remember picture,overlay]
\node[text width = 7in, align=center] at (current page text area.south) {\textcopyright 2020 California Institute of Technology. Government sponsorship acknowledged};
\end{tikzpicture}

 \begin{tikzpicture}[remember picture,overlay]
\node[text width = 8in, align=center] at ([yshift=1cm] current page text area.north) {\textcolor{black}{Accepted document version.\\ AIAA Journal of Spacecraft \& Rockets (JSR) 2020, Vol. 57 No. 4. DOI: https://doi.org/10.2514/1.A34437}};
\end{tikzpicture}


\section*{Nomenclature}


{\renewcommand\arraystretch{1.0}
\noindent
\begin{tabular}{l @{\quad=\quad} l}
		$a$ &  speed of sound [m/s] \\
		$C_D$& drag coefficient \\
		$d$   & diameter of probe [m] \\
		$E$   & Young's modulus [Pa] \\
		$F_D$   & drag force [N] \\
		$f$   & factor of safety $(f\geq1)$ \\
		$g$   & local gravitational acceleration [m/s\textsuperscript{2}] \\
		$h$ & thickness [m] \\
		$\Delta h$ & heat of fusion [J/kg] \\
		$k$   & conductivity [W/m-°C] \\
		$P$   & pressure [Pa] \\
		$\Pra$  & Prandtl number \\
		$M$  & Mach number \\
		$m$  & mass [kg] \\
		$\Nu$ & Nusselt number \\
		$\dot{Q}$ & heat flow rate [W] \\
		$r$   & radius of probe [m] $(r=d/2)$ \\
		$R$   & thermal resistance [°C/W] \\
		$\Rey$ & Reynolds number \\
		$T$   & temperature [°C]  \\
		$t$   & time [s]  \\
		$\boldsymbol{v}$   & velocity [m/s] \\
		$\mathcal{V}$ & volume [m\textsuperscript{3}] \\
		$z$  & altitude [m] \\
		$\beta$ &  ballistic coefficient [kg/m\textsuperscript{2}] \\
		$\nu$ & kinematic viscosity [m\textsuperscript{2}/s] \\
		$\eta$ & penalization factor $(\eta\leq1)$\\
		$\rho$ & density [kg/m\textsuperscript{3}]\\
	\end{tabular}}



\section{Introduction}
\lettrine{V}{ehicles} that explore the lower atmosphere of Venus must be designed to tolerate the high temperature and high pressure environment that reaches 462°C and 92 bar at the surface \cite{seiff1985models}. All past missions have used insulated pressure vessels to protect the payload, providing it with a benign environment for the short time needed to reach the surface and make scientific measurements along the way \cite{huntress2011soviet,bienstock2004pioneer}. This approach has been very successful, but has resulted in relatively large vehicles ranging from 99 kg  \cite{bienstock2004pioneer} for the Pioneer Venus (PV) small probe to 716 kg \cite{huntress2011soviet} for the VEGA-2 lander. The purpose of the study reported here is to determine from a thermo-mechanical preliminary design perspective what the lower mass limit is for this kind of short-duration probe that employs the insulated pressure vessel architecture. This miniaturization is motivated by potential future mission opportunities for adding one or more probes as secondary payloads on a spacecraft going to Venus, and clearly smaller probes are more readily accommodated than larger ones. Continuing miniaturization and improvement of science instrumentation also facilitate the use of small but capable probes that carry less payload mass than their PV and Venera-VEGA predecessors.

Small Venus probes have the advantage that less heat flows into the payload because of the reduced surface area. However, the ability to absorb that heat without exceeding an allowable operating temperature is also reduced given the smaller mass available. The challenge for miniaturization is that surface area scales with the radius squared but the thermal absorption mass scales with the radius cubed, and this must place a fundamental limit on how small the probe can be and still maintain a survivable operating temperature all the way to the surface.  For example, the VEVA mission proposal \cite{klaasen2003veva} envisioned four small 3.5 kg imaging probes dropped by a balloon \cite{kerzhanovich2003low}. The study of \citet{lorenz1998design} concluded that even smaller probes on the order of 1 kg are possible but only if the payload is ruggedized to tolerate an elevated temperature of 100°C and full Venus pressure of 92 bar. Extremely small probes, on order of 100 g, are discussed in an ESA microprobe design by \citet{wells2004atmospheric} if the minimum altitude requirement is relaxed to 30 km \cite{wells2004atmospheric} or 10 km \cite{vandenberg2005esa} rather than to the surface.

In this study, we examine the case where full temperature and pressure protection of the payload is provided all the way to the surface. In addition to traditional bluff-body probe designs, we also determine relationships for how streamlining the probe can further minimize the insulation mass required, at the cost of reducing the time-to-surface for data collection and transmission.

The paper is organized as follows: First we describe a simple drop probe model that captures the relevant physics while enabling extensive trade studies. Second, we perform a parametric study on a variety of probe designs and derive the scaling laws for the thermal and pressure vessel subsystems. Finally, we invert these scaling laws to address the design problem, determining a minimum mass cut-off for probes and discussing the associated trade space.

\section{Methodology \label{sec:methods}}
\subsection{Probe Model}
Venus entry is generally performed with a heat shield that is ejected after use, allowing the spacecraft to shed both mass and absorbed entry heat before plunging into the hot lower atmosphere. All Venus entries have been performed in this manner, with the partial exception of the Pioneer Venus Small Probe, which descended with the heat shield still attached \cite{bienstock2004pioneer}. Our focus for this paper is on probes, not landers, so no attempt will be made to cushion the landing or live on the surface for appreciable time (although one of the PV small probes did survive its landing without being specifically designed to do so \cite{bienstock2004pioneer}).

\begin{figure}[t]
	\centering
	\includegraphics[width=\figwidth]{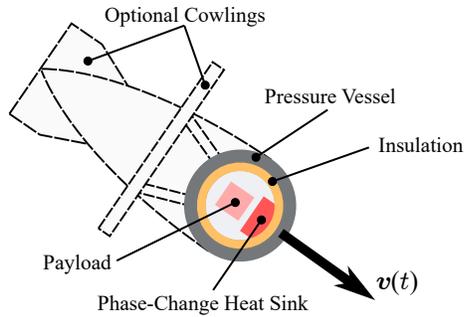}
	\caption{Descent Probe Schematic \label{fig:probeCartoon}}
\end{figure}

Our model, for the sake of consistency, assumes that entry into the atmosphere has been completed and heat shield ejected, leaving only the probe. Accordingly, our model furthermore applies to a probe dropped from a high-altitude Venus aerial platform. This probe, illustrated in Fig. \ref{fig:probeCartoon}, consists of a layered construction of five components in analog of Pioneer Venus Probes and the Venera/Vega landers:
\begin{enumerate}
	\item A spherical pressure vessel made of titanium alloy Ti-6Al-4V.
	\item An optional external cowling (assumed negligible mass), designed to reduce or augment the drag during descent. Such aerodynamic devices are generally only a few percent of the total mass for mature mass-breakdowns \cite{huntress2011soviet}, and different types (plates, tailboxes, or otherwise) have been shown to package into entry systems in tandem with other assets \cite{klaasen2003veva,vandenberg2005esa,huntress2011soviet}, so we make the simplifying assumption to ignore this small mass in this study.
	\item An insulation layer, placed inside the pressure vessel similar to the PV probes. Our selected material is a calcium-silicate insulation (ZIRCAL-18 \cite{zrci2018zircal}) suitable for high-temperature application, though the PV probes instead used MLI blankets. Fiberglass filled with xenon gas \cite{hall2000thermal} is another proposed insulator. 
	\item A phase-change heat sink material, composed of a lithium salt (lithium nitrate trihydrate LiNO\textsubscript{3}-3H\textsubscript{2}O) that melts at $30^\circ$C, technology developed for the Venera/Vega landers \cite{huntress2011soviet}.
	\item A payload, which claims the remaining volume and mass. The payload and heat-sink material are assumed to share the spherical volume inside the insulation.
\end{enumerate}

\begin{table*}[tb]
	\centering
	\begin{tabular}{llcccc}
		\hline \hline
		Material & Use & $\rho$ [g/cm\textsuperscript{3}]& $E$ [GPa] & $k$ [W/m-°C] & $\Delta h$ [kJ/kg] \\\hline
		Titanium 6Al-V4 & Pressure Vessel & 4.43 & $113$ & 7 & --\\
		ZIRCAL-18& Insulation & 0.280 & -- & 0.07 & -- \\
		LiNO\textsubscript{3}-3H\textsubscript{2}O& Heat Sink & 1.50 & -- & -- & 296\\
		\hline
	\end{tabular}
	\caption{\label{tab:materials} Selected Material Properties of Probe Components}
\end{table*}

The relevant material properties of these components are listed in Table \ref{tab:materials}. Future material science developments will naturally trickle into improved properties, but the current selection provides a reasonable baseline for mission design.

\subsection{Descent Model \label{sub:descent}}
Our descent model begins at $z=65$ km altitude above the surface of Venus, starting from an assumed initial velocity of $||\mathbf{v}(0)||=200$ m/s oriented $30^\circ$ below the horizon, similar to the Vega 2 flightpath \cite{huntress2011soviet}. Including a hypersonic entry above 65 km does not affect the results of our simulation, as the heat shield absorbs all the entry heat and then detaches.

As the probe descends, we interpolate the local atmospheric properties $\rho_\text{atm}$, $\nu_\text{atm}$, $a_\text{atm}$, $T_\text{atm}$, $P_\text{atm}$, and $k_\text{atm}$ (density, kinematic viscosity, speed of sound, temperature, pressure, and thermal conductivity) from the equatorial Venus International Reference Atmosphere (VIRA) \cite{seiff1985models}, thereby determining the instantaneous Reynolds ($\Rey$) and Mach numbers ($M$):
\begin{equation}
\label{eq:ReMach}
\Rey(t) = \frac{||\boldsymbol{v}(t)|| \, d}{\nu_\text{atm}(z)}; \,\,\,\,\, M(t) = \frac{||\boldsymbol{v}(t)||}{a_\text{atm}(z)}
\end{equation} 
where $d=2r$ is the diameter of the probe. For any body, the drag coefficient $C_D$ is a function of the instantaneous $\Rey$ and $M$, meaning a $C_D$ lookup table can be used to determine the drag force $F_D(t)$:
\begin{align}
C_D(t) &= C_D(\Rey(t),M(t)) \label{eq:Cd}\\
F_D(t) &= \frac{1}{2}\rho_\text{atm}(z)||\boldsymbol{v}(t)||^2 \pi r^2 C_D(t)  \label{eq:Fd}
\end{align}

For all of the probes analyzed, the Reynolds number varies little with time and stays well above natural turbulent transition for external boundary layers, generally around $\Rey\approx10^6$ (see Sec \ref{sec:results}.\ref{sub:trajectoryResults}), and most bodies transition earlier \cite{hoerner1965fluid}. In this regime, drag for both bluff and streamlined objects is only weakly dependent on Reynolds number, generally at or less than a scaling of $\Rey^{-1/6}$ \cite{hoerner1965fluid}. This allows for the simplifying assumption in the analysis below that the drag coefficient is independent of Reynolds number.

Accordingly, we can invert the drag problem and consider the \emph{$C_D$ as a design input} that the cowling is assumed to match once subsonic speeds are reached, limited by values plausible within this flow regime ($\Rey>10^6$, $M<0.8$).  All landers and probes dropped to the Venus surface to-date have had drag coefficients higher than that of the pressure vessel sphere itself, either through the addition of a drag plate (Venera/Vega), a heat shield (PV-SP), or a large lip with spin vanes (PV-LP). However, a wide range of designed drag coefficients are possible (Fig. \ref{fig:probeCartoon}), for example:
\begin{enumerate}
	\item A drag plate with roughly twice the diameter of the probe, bringing the drag coefficient up to $C_D=4$ normalized to the pressure vessel frontal area. To capture higher Mach numbers, we assume the expression rounded from \citet{hoerner1965fluid} to model the relatively brief supersonic descent phase:
	\begin{align}
	C_{D,\text{plate}}(M) &= 1.5+2.5(1+0.25M^2) \label{eq:Cdplate} 
	\end{align}
	\item A sphere with mild surface roughness, leading to a streamlined $C_D=0.2$. Above $M=0.8$, we assume a linear drag rise to $C_D(M)=0.9$ \cite{hoerner1965fluid}.
	\begin{align}
	[M \leq 0.8]: \,\, C_{D,\text{sphere}}(M) &= 0.2 \label{eq:Cdsphere} \\
	[0.8 < M < 1]: \,\, C_{D,\text{sphere}}(M) &= 0.2+3.5(M-0.8) \\
	[M \geq 1]: \,\, C_{D,\text{sphere}}(M) &= 0.9 
	\end{align}
	\item A tailbox, dropping the drag coefficient to $C_D=0.05$, the approximate minimum configuration reported by \citet{hoerner1965fluid}. Above $M>0.8$, we again assume a linear drag rise to $C_D(M)=0.9$, similar to the sphere.
	\begin{align}
	[M \leq 0.8]: \,\, C_{D,\text{tailbox}}(M) &= 0.05 \\
	[0.8 < M < 1]: \,\, C_{D,\text{tailbox}}(M) &= 0.05+4.25(M-0.8) \\
	[M \geq 1]: \,\, C_{D,\text{tailbox}}(M) &= 0.9 \label{eq:Cdtailbox} 
	\end{align}
\end{enumerate}

Note that all of these bodies would likely require spin-stabilization for the first few minutes of entry \cite{lorenz2007spinning} during the supersonic phase. The equation of motion of the probe, including both gravity and buoyancy, is then:
\begin{equation}
\label{eq:eqmotion}
m\dot{\boldsymbol{v}} = -\hat{\boldsymbol{v}}(t) \cdot F_D(t) \,\,\, -\hat{\boldsymbol{z}} \cdot m g(z) \,\,\, -\hat{\boldsymbol{z}} \cdot \frac{4}{3} \pi r^3 \rho_\text{atm}(z) g(z)
\end{equation}
where $\hat{\boldsymbol{v}}$ and $\hat{\boldsymbol{z}}$ are the unit vectors in the direction of $\boldsymbol{v}$ and $\boldsymbol{z}$, $m$ is the mass, and $g(z)$ is the local Venus gravitational acceleration from the VIRA model \cite{seiff1985models}:
\begin{equation}
g(z) = G \frac{M_\text{Venus}}{(z+r_\text{Venus})^2}
\end{equation}
assuming a mean Venus radius of $r_\text{Venus} = 6052$km, mass of $M_\text{Venus}=4.867\times10^{24}$kg, and universal gravitational constant $G=6.674\times10^{-11}$. All equations of motion are integrated in Matlab, utilizing the ode45 Runge-Kutta numerical solution \cite{dormand1980family}.

\subsection{Thermal Model}
The thermal model assumes a steady-state heat flow rate is obtained at every time instant during the descent, consisting of four steps: (1) convection from the atmosphere to the skin of the probe, (2) conduction through the pressure vessel, (3) conduction through the insulation, and finally (4) absorption into the phase-change heat sink. The thermal network (Fig. \ref{fig:thermalnetwork}) therefore comprises of a set of series resistances ($R_\text{convect}$, $R_\text{vessel}$, and $R_\text{insul}$) to determine the net heat flux:

\begin{figure}[t]
	\centering
	\includegraphics[width=\figwidth]{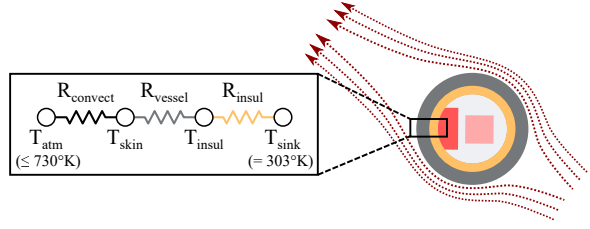}
	\caption{Thermal Network Model \label{fig:thermalnetwork}}
\end{figure}

\subsubsection{Convection from the Atmosphere}
We model the convection problem as a simple dropping sphere, assuming the downstream cowling structural attachments are designed to have a negligible heat path compared to the stagnation flow on the probe front.  At subsonic speeds, \citet{achenbach1978heat} provides heat convection correlations (Nusselt $\Nu$ as a function of $\Rey$) for spheres in a gas of Prandtl number $\Pra=0.71$.  Also available from the VIRA data, $\Pra(z)=0.71$ is a faithful approximation for Venus atmosphere due to the majority carbon dioxide constituency. 
\begin{align}
\label{eq:achenbach}
[\Rey_\text{film}<2\times 10^5]:& \,\,\,\,
 \Nu \approx 2+\left(0.25\Rey_\text{film}+3\times 10^{-4}\Rey_\text{film}^{8/5}\right)^{1/2} \\
[4\times 10^5<\Rey_\text{film}<5\times 10^6]:& \,\,\,\,
\Nu \approx 430+5\times 10^{-3}\Rey_\text{film} + ... \,  \text{\small [smaller high-order terms]}
\end{align}

\citet{achenbach1978heat} claims the Reynolds exponent drops back again to $\sfrac{4}{5}$ beyond $\Rey_\text{film}>5\times 10^6$, but does not give another relation for this regime. At the risk of slight overestimation of the convective coefficient and to retain the simplest possible model, we keep his linear correlation beyond $\Rey_\text{film}>5\times 10^6$. In Achenbach's relation, the Reynolds number must also be corrected for the new gas properties of the lower-temperature ``film'' near the probe, where $T_\text{film}$ is generally taken as the average of the atmospheric temperature and probe surface temperature. We assume a CO\textsubscript{2}-like correction for the kinematic viscosity, with properties from \citet{fenghour1998viscosity}, as CO\textsubscript{2} is by far the dominant atmospheric constituent (96.5\% in the VIRA model \cite{seiff1985models}):
\begin{align}
\label{eq:Refilm}
\Rey_\text{film}(t) = \Rey(t) \frac{\nu_\text{CO2}(T_\text{atm})}{\nu_\text{CO2}(T_\text{film})}
\end{align}

Finally, we can use  Achenbach's relation to approximate the convective resistance $R_\text{convect}$ from the atmospheric temperature to the pressure vessel skin:
\begin{equation}
\label{eq:CdNu}
R_\text{convect}(t) = \frac{d}{4\pi r^2 k_\text{atm}(z)\, \Nu(Re_\text{film})}
\end{equation}

\subsubsection{Conduction through the Pressure Vessel}
We determine the required pressure vessel thickness $h_\text{vessel}$ from \citet{young2002roark} spherical buckling criterion for the maximum atmospheric pressure $P_\text{atm}(0)$, further including a factor-of-safety of $f_\text{vessel} = 1.3$ and Young's modulus weakened for high-temperature at $\eta_\text{T}=0.76$ \cite{handbook1998mil}.  
\begin{equation}
\label{eq:pressureVessel}
\frac{h_\text{vessel}}{r}  = \sqrt{\frac{f_\text{vessel} P_\text{atm}(0)}{0.365 \eta_\text{T} E_\text{vessel}}}
\end{equation}

The titanium pressure vessel does not provide any practical insulation against the rising external temperature, but its conductive resistance can be easily computed from spherical shell relations:
\begin{equation}
\label{eq:sphereResistance}
R_\text{vessel} = \frac{1}{4\pi k_\text{vessel} } \frac{h_\text{vessel}}{r(r-h_\text{vessel})}
\end{equation}

\subsubsection{Conduction through the Insulation}
The series conductive resistance of the insulation layer $R_\text{insul}$ beneath is similarly computed, with radius $r_\text{insul}=r-h_\text{vessel}$. As a rough model for heat leaks from sensor pass-throughs, we penalize the insulation effectiveness by $\eta_\text{insul} = 0.5$. 
\begin{equation}
\label{eq:insulResistance}
R_\text{insul} = \frac{\eta_\text{insul} }{4\pi k_\text{insul} } \frac{h_\text{insul}}{r_\text{insul}(r_\text{insul}-h_\text{insul})}
\end{equation}

The insulation effectiveness $\eta_\text{insul}$ may well scale with probe size and type of science instruments selected, but is functionally equivalent to an effective change in insulation effectiveness (discussed in Sec. \ref{sec:results}.\ref{sub:material}).

\subsubsection{Net Heat Flux}
The external heat $\dot{Q}_\text{external}$ conducted through the layers is therefore determined from the series resistances and the heat sink melting temperature of $T_\text{sink}=30^\circ$C.
\begin{gather}
\label{eq:seriesResistance}
\dot{Q}_\text{external} = \frac{T_\text{atm}(z)-T_\text{sink}}{R_\text{convect}+R_\text{vessel}+R_\text{insul}} \\
\label{eq:melting}
\dot{m}_\text{sink,melted} = \frac{\dot{Q}_\text{external}+\dot{Q}_\text{internal}}{\Delta h_\text{sink}}
\end{gather}
where $\Delta h_\text{sink}$ is the specific heat of fusion of the heat sink material, $\dot{m}_\text{sink,melted}$ is the rate of melting in kilograms per second, and $\dot{Q}_\text{internal}$ covers the heat generation from internal to the insulation. We assume $50$ W of internal heat estimated from the PV-SP design: 10 W of instrument power \cite{bienstock2004pioneer}, a 10 W transmitter \cite{bienstock2004pioneer} operating at a typical 25\% efficiency to dissipate 30 W of heat, and a final $10$ W for the computer and power management system. These $50$ watts increase the rate of lithium melting by an extra 10 g/min, so designs with other internal power levels can be scaled accordingly. 

For all trajectories, we simulate the initial temperature of the probe at $30^\circ$C, the melting point of the lithium salt heat sink; this assumes that the heatshield includes a thermal solution to provide this initial condition after entry. Subsequently, we then assume that all additional heat is absorbed by the phase-change heat sink material, thereby keeping the payload at constant temperature for the entire descent. This is a conservative assumption made for computational simplicity. Inclusion of the non-salt payload's ability to absorb heat will reduce the amount of salt required. Assuming a mostly metallic payload that is allowed to heat by 50$^\circ$C, each gram of salt absorbs roughly 12 times more energy than a gram of titanium (\textasciitilde500 J/kg-$^\circ$C) and 6.6 times more than aluminum (\textasciitilde900 J/kg-$^\circ$C); meaning our thermal capacity can be conservative by up to a factor of two for probes of large payload and little salt (see Section \ref{sec:results}:\ref{sub:params}).

We also however, include a thermal factor-of-safety $f_\text{sink}=1.3$ on the carried heat sink material, so 30\% extra remains on impact if the predicted uniform heat transfer is realized. This safety factor is meant to cover any ``hot spots'', where some extra reservoirs of the salt may needed to compensate for trouble areas in a complex interior. We also assume $f_\text{sink}$ will cover uncertainties in the atmospheric properties and flow characteristics around the probe (a conservative estimate given only mild density and temperature fluctuations from PV data \cite{seiff1980measurements} and a largely predictable subsonic descent). 

The bonus heat capacity of the pressure vessel may be ignored as it resides outside the insulation layer: the convective resistance is much lower than the conductive (see Section \ref{sec:results}.\ref{sub:thermalResults}) so this capacity is quickly saturated by the rapid heat transfer. The Venera/Vega landers had insulation on both sides of the pressure vessel \cite{huntress2011soviet} which allowed use of the vessel heat capacity, but this architecture is generally not used in smaller probes \cite{wells2004atmospheric,klaasen2003veva,bienstock2004pioneer} that are more volume-constrained.

\section{Results \label{sec:results}}
\subsection{Parameterization \label{sub:params}}
\begin{figure}[t]
	\centering
	\includegraphics[width=\figwidth]{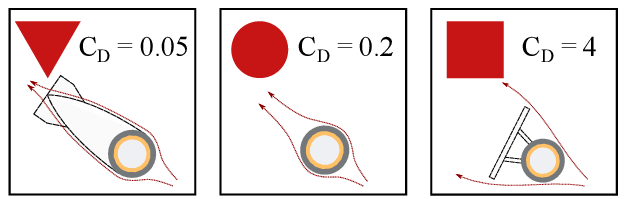}
	\includegraphics[width=\figwidth]{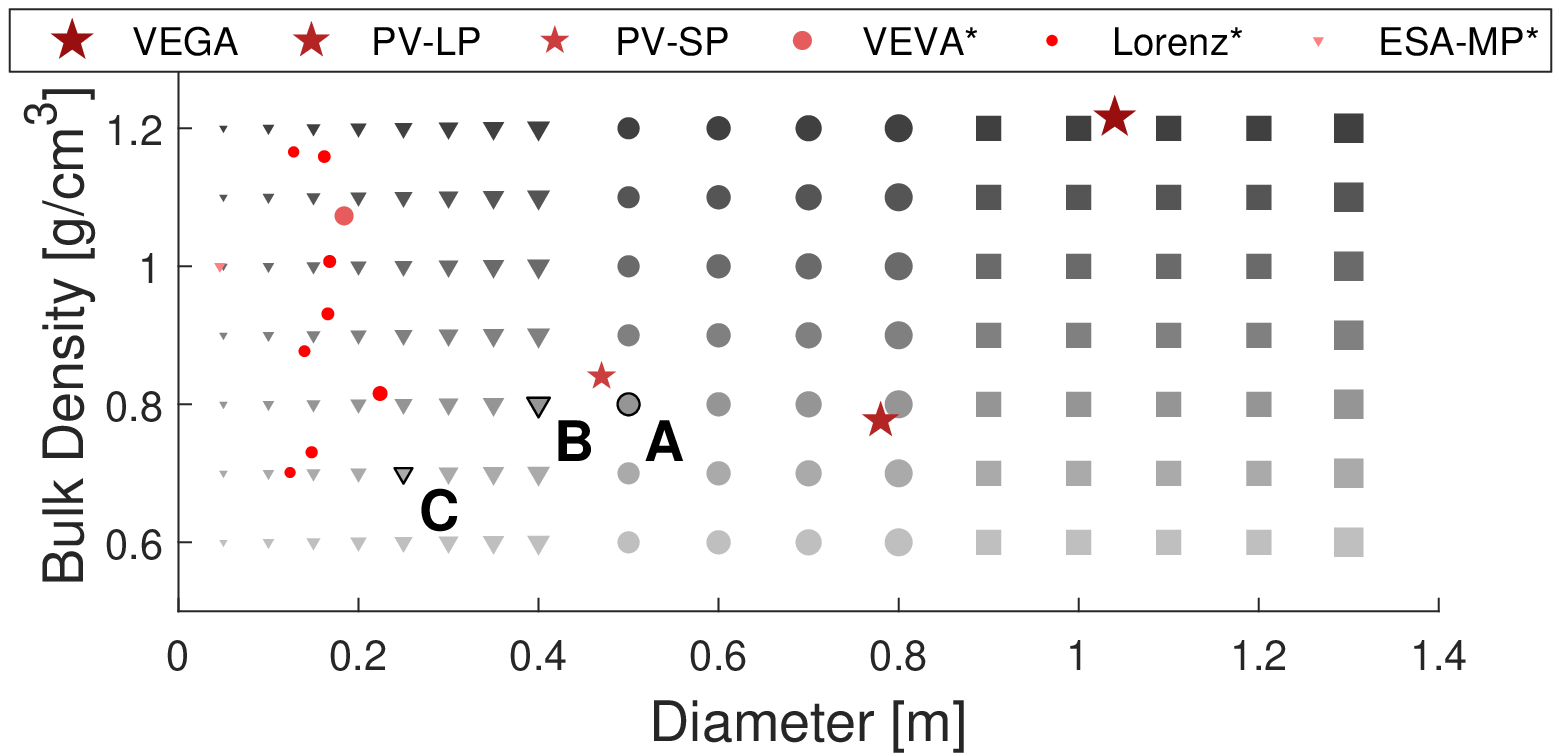}
	\caption{Simulated Probe Design Space \label{fig:dataKey}}
\end{figure}

The full range of simulated probes is enumerated in Fig. \ref{fig:dataKey}; this figure can be used as a data key for the remainder of this paper. Our parameters are selected to roughly overlap a wide array of probes and landers either already flown to Venus or detailed design concepts (marked with asterisk *). The diameter $d$ of these probes varies from $0.1$ to $1.3$ m, and bulk density $\rho_\text{bulk}$ (including the pressure vessel and everything inside it) varies from $0.6$ to $1.2$ g/cm\textsuperscript{3}. While certainly low compared to metallic materials, the bulk density of probes is usually limited to $\rho_\text{bulk} \leq 1.2$ due to volume packing constraints. Mass can be derived from the chosen bulk density and radius as $m= \rho_\text{bulk}\mathcal{V} = \frac{4}{3}\rho_\text{bulk}\pi r^3$. VEGA parameters are taken from \citet{huntress2011soviet}, PV-LP/PV-SP from \cite{,bienstock2004pioneer}, VEVA diameter and mass from \citet{klaasen2003veva} and \citet{kerzhanovich2003low} respectively, Lorenz from \citet{lorenz1998design}, and ESA Microprobe from \citet{wells2004atmospheric}. 

In Fig. \ref{fig:dataKey} and throughout the remainder of the paper, each icon is scaled in size with the diameter of the selected probe, and its shading is a function of the bulk density. Icons shape gives the probe drag characteristics discussed in Sec. \ref{sec:methods}:\ref{sub:descent} - square symbols represent drag plates, circles represent spheres, and triangles represent streamlined probes. For the majority of the descent these shapes have the subsonic drag coefficients of $C_D=4$, $C_D=0.2$, and $C_D=0.05$ respectively, but during the brief supersonic phase variable drag coefficients are used instead (Eq. \ref{eq:Cdplate} - \ref{eq:Cdtailbox}).

We also specifically call out a few design points (A, B, etc.) enumerated in Table \ref{tab:probes} for detailed analysis later in the manuscript to showcase results. For example, Probe A is spherical with drag coefficient $C_D=0.2$, has diameter $d=0.5$ m, and a bulk density of $\rho_\text{bulk}=0.8$ g/cm\textsuperscript{3}; comparable in diameter and mass to the Pioneer Venus Small Probe pressure vessel, though more streamlined. The smallest probes, such as Probe C, carry very little heat sink material and the 10 g/min melt due to internal heating ($m_\text{sink,internal}$) becomes a significant fraction of their thermal load.

\begin{table*}[tb]
	\centering
	\begin{tabular}{l | ccc | cc | ccccccc}
		\hline \hline
		\multirow{ 2}{*}{Id.} & $d$ & $\rho_\text{bulk}$ & $C_D$ & $m$ & $\beta$ & $h_\text{vessel}$ & $h_\text{insul}$ & $m_\text{sink}$ & $m_\text{sink,internal}$ & $m_\text{payload}$ & $\rho_\text{payload}$ & $t_\text{descent}$ \\ 
		& [m] & [g/cm\textsuperscript{3}] & -- & [kg] & [kg/m\textsuperscript{2}] & [mm] & [mm] & [kg] & [\%] & [kg] & [g/cm\textsuperscript{3}] & [min] \\ \hline
		A & 0.50 & 0.80 & 0.2 & 52 & 1300 & 4.9 & 34 & 4.8 & 5 & 24 & 0.68 & 23 \\
		B & 0.40 & 0.80 & 0.05 & 27 & 4300 & 3.9 & 25 & 2.4 & 5 & 13 & 0.67 & 13 \\
		C & 0.25 & 0.70 & 0.05 & 5.7 & 2300 & 2.4 & 37 & 0.91 & 20 & 1.3 & 0.64 & 17 \\
		\hline
	\end{tabular}
	\caption{\label{tab:probes} Selected Example Probe Parameters}
\end{table*}

Again, none of the tested probes include parachutes, similar to the PV Small Probe. In order to make a comparison from our results to temporarily parachuted Venus platforms (such as Venera/Vega and the PV Large Probe), all reported descent times are measured from 50 km rather than the initial 65 km, as parachute jettison is generally performed at that altitude \cite{huntress2011soviet,bienstock2004pioneer}. 

\subsection{Trajectory Correlations \label{sub:trajectoryResults}}
We first note that the lower drag coefficients shorten the descent timeline, as compared to the fully separated flow on classic Venus probes. The heaviest computed streamlined probes will reach the surface on the order of 10 to 20 minutes, rather than 40 to 60 minutes. As is common for falling objects in a set atmosphere and gravity, the trajectory details are solely dependent on the ballistic coefficient $\beta$.
\begin{equation}
\label{eq:ballistic}
\beta = \frac{m}{\pi r^2 C_D}
\end{equation}

\begin{figure}[t]
	\centering
	\includegraphics[width=\figwidth]{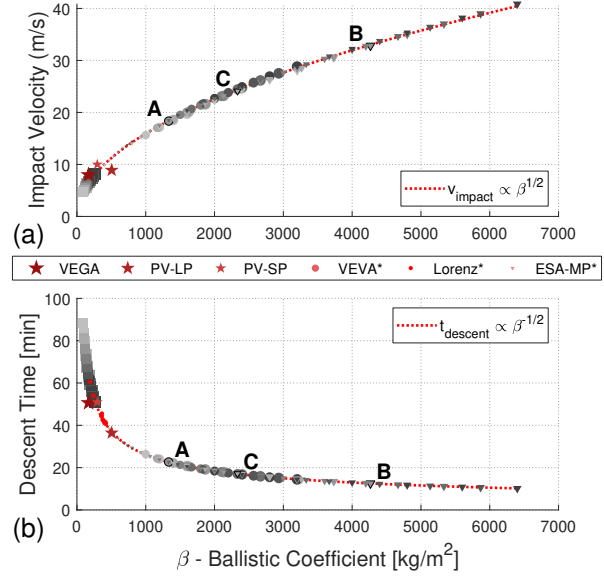}
	\caption{Descent Trajectory Correlations \label{fig:trajCorrelate}}
\end{figure}
\begin{figure}[t]
	\centering
	\includegraphics[width=\figwidth]{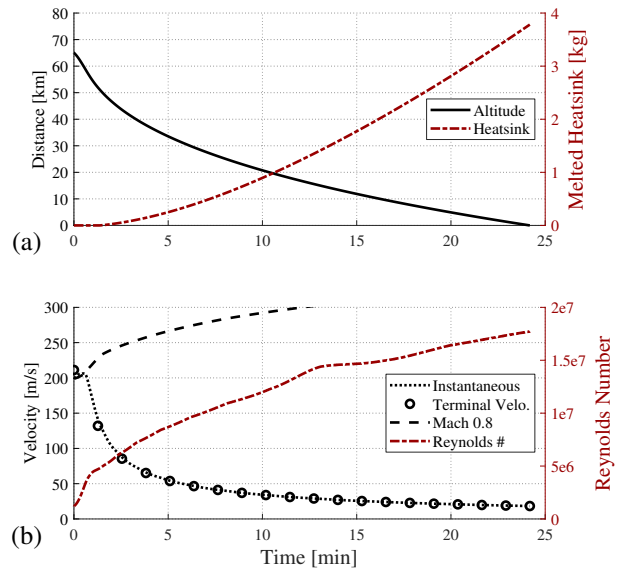}
	\caption{Descent Timeseries (Probe A)  \label{fig:flightTrajectory}}
\end{figure}

For a probe at terminal velocity, we can explicitly derive that the impact velocity is proportional to $\beta^{1/2}$ and descent time as $\beta^{-1/2}$, consistent with Fig. \ref{fig:trajCorrelate}:

\begin{align}
v_z|_{z=0} = \hat{\boldsymbol{z}}\cdot\mathbf{v}(z)|_{z=0} &= \left. \sqrt{\frac{2mg(z)}{\rho_\text{atm}(z)\pi r^2 C_D}} \, \right|_{z=0} \propto \beta^{1/2} \label{eq:vz} \\
t_\text{descent} &=  \int_{0}^{z(t=0)} \frac{1}{v_z(\zeta)} d\zeta  \propto \beta^{-1/2} 
\end{align}

These relations assume that the effect of buoyancy is relatively small, so are not expected to hold as the bulk density approaches the atmospheric density. Given the streamlining, the ballistic coefficients investigated can be well over 1000 kg/m\textsuperscript{2}.

The tight agreement in Fig. \ref{fig:trajCorrelate} illustrates that the probe quickly reaches terminal velocity for the given altitude, meaning that the impact velocity and descent time are dependent on the probe's ballistic characteristics, rather than the Venus entry parameters. Consequently, if the probe were dropped instead from an aerial platform at much lower initial velocity, the trajectory would be largely equivalent. Furthermore, the tight agreement in Fig. \ref{fig:trajCorrelate} also shows that Mach effects have little influence on the trajectory past 50 km, as only the subsonic drag coefficient is used in our ballistic coefficient in Eq. \ref{eq:ballistic}. For example, if we investigate the specific trajectory from Probe A (Fig. \ref{fig:flightTrajectory}a), we see that terminal velocity is reached and Mach number falls below subsonic within the first minutes of descent (Fig. \ref{fig:flightTrajectory}b). All reasonable initial velocity estimates provide similarly short timescales to terminal velocity and subsonic flight.

\begin{figure}[t]
	\centering
	\includegraphics[width=\figwidth]{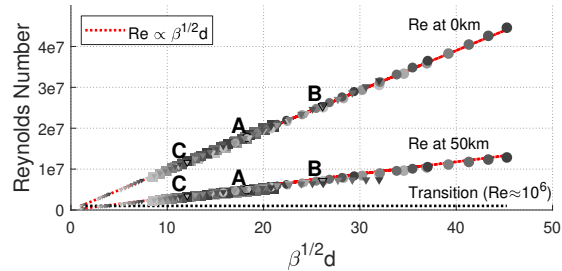}
	\caption{Vessel Diameter Reynolds Number at two Altitudes \label{fig:ReCorrelate}}
\end{figure}

The Reynolds number ramps up surprisingly slowly once the probe passes 50 km altitude (Fig. \ref{fig:flightTrajectory}b), growing by less than a factor of four for each probe tested and staying above turbulent transition ($\Rey>10^6$ \cite{hoerner1965fluid}) for external flow boundary layers (Fig. \ref{fig:ReCorrelate}).  This phenomenon is due to the opposite scaling of atmospheric density and kinematic viscosity with altitude as the probe falls, and furthermore justifies modeling approximations of a single drag coefficient for the descent. 
At any given altitude, $\Rey$ scales with the velocity and diameter of the vessel:
\begin{equation}
\label{eq:ReBeta}
\Rey (z) = \frac{d||\boldsymbol{v}(z)||}{\nu_\text{atmo}(z)} \propto \frac{\beta^{1/2}d}{\nu_\text{atmo}(z)}
\end{equation}

Combining Eq. \ref{eq:vz} and \ref{eq:ReBeta}, we note that in a hypothetical atmosphere where $\nu_\text{atm}(z) \propto [\rho_\text{atm}(z)]^{-0.5}$, an object falling at terminal velocity will stay at a perfectly fixed Reynolds number for the entire descent. The VIRA model \cite{seiff1985models} predicts a relationship closer to $\nu_\text{atm} \propto \rho_\text{atm}^{-0.9}$, which accounts for the slow Reynolds number growth.
 

\subsection{Thermal Correlations \label{sub:thermalResults}}
Choosing a specific drag coefficient clearly affects the thermal design: a faster dropping probe will have stronger external convection, while a slower probe will have a longer timeline for internal conduction to occur through the insulation. Depending on the parameters of the problem, one would expect that dropping faster or slower are both viable strategies to minimize thermal effects. 

\begin{figure}[t]
	\centering
	\includegraphics[width=\figwidth]{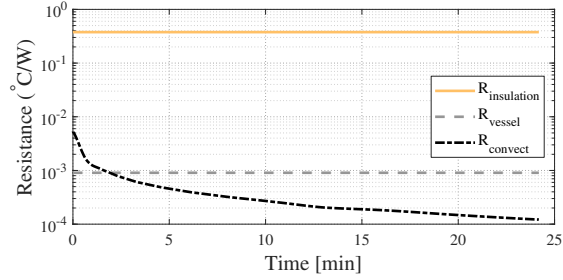}
	\caption{Thermal Resistance in Descent (Probe A) \label{fig:thermalPerformance}}
\end{figure}

However, the particulars of the Venus environment make \emph{dropping faster always more thermally efficient} once entry is performed, as the thermal resistances act in series and the insulation is by far most limiting, and supersonic heating of the Venus gas is negligible after heat shield jettison. Taking Probe A again as an example in Fig. \ref{fig:thermalPerformance}, we note that the convection and pressure vessel are of similar resistance to the heat flow, while the insulation required to ensure survival is 2.5 orders of magnitude more resistant ($R_\text{insul} \gg R_\text{convect}$ and $R_\text{insul} \gg R_\text{vessel}$).  \citet{achenbach1978heat} predicts a roughly linear relationship between the airspeed and convective transfer coefficient, meaning that the probe has to drop orders of magnitude slower before the convective resistance need be considered as anything other than zero. In other words, convection is so strong that the skin temperature rapidly saturates at the local atmospheric temperature, and the insulation then constricts the heat flow from there. 

Furthermore, the extremely high convection rate reduces our analysis sensitivity to Achenbach's heat transfer correlations. An infinite convection coefficient would act similarly to Achenbach's - the insulation is the only effective resistive bottleneck for the heat flow.

\begin{figure}[t]
	\centering
	\includegraphics[width=\figwidth]{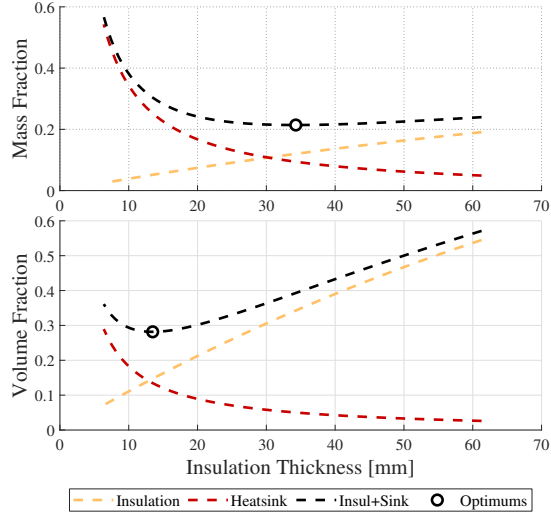}
	\caption{Optimized Insulation Thickness (Probe A) \label{fig:thermalOptimum}}
\end{figure}

Applying this observation to the design problem, a spectrum of solutions exist in each probe for allocating mass between the insulation layer and the heat sink. Using more heat sink material reduces the required insulation and vice versa, although the heat sink mass rises rapidly with very thin insulation. Figure \ref{fig:thermalOptimum} illustrates the tradeoff between the two components as the insulation thickness is varied in Probe A.

An additional constraint is needed to close the problem, such as a cost function to minimize. Of note, both the minimum mass and minimum volume optimizations (shown in Fig. \ref{fig:thermalOptimum}) can be solved analytically for the dominant conduction effect. Referring first to the minimum mass solution, the total mass of the thermal components can be found by manipulating Eqs. \ref{eq:sphereResistance} - \ref{eq:melting}:
\begin{align}
\label{eq:massInsulAndSink}
m_\text{insul} &= \frac{4}{3}\pi\rho_\text{insul} \left[r_\text{insul}^3-(r_\text{insul}-h_\text{insul})^3\right] \\
m_\text{sink} &= \int_{0}^{t_\text{descent}} \frac{T_\text{atm}(\tau)-T_\text{sink}}{\Delta h_\text{sink} \, R_\text{insul}} d\tau \\ 
&+ \frac{t_\text{descent}\dot{Q}_\text{internal}}{\Delta h_\text{sink}} \nonumber
\end{align}

Performing the minimization of the net mass including safety factors ($m_\text{insul}+f_\text{sink} m_\text{sink}$) with respect to the insulation thickness we obtain:
\begin{equation}
\label{eq:calculus}
\frac{\partial m_\text{insul}}{\partial h_\text{insul}} + \frac{\partial m_\text{sink}}{\partial h_\text{insul}} = 0 , \,\,\, \frac{\partial^2 m_\text{insul}}{\partial h^2_\text{insul}} + \frac{\partial^2 m_\text{sink}}{\partial h^2_\text{insul}} > 0
\end{equation}
\begin{equation}
\label{eq:optim}
h_\text{insul,optim} = \frac{1}{2}r_\text{insul}-\frac{1}{2}r_\text{insul}\sqrt{1-4\sqrt{K/r_\text{insul}^2}}
\end{equation}

where quantity $K$ is dependent on the probe geometry and trajectory as:
\begin{equation}
\label{eq:C}
K = \frac{f_\text{sink}\eta_\text{insul}k_\text{insul}\int_{0}^{t_\text{descent}} [T_\text{atm}(\tau)-T_\text{sink}] d\tau}{\Delta h_\text{sink} \rho_\text{insul}}
\end{equation}

In the thin shell limit ($h_\text{insul}/r_\text{insul} \ll 1$), Eq. \ref{eq:optim} reduces to $h_\text{insul,optim} \approx \sqrt{K}$. Replacing $\rho_\text{insul}$ with $\rho_\text{sink}$ in Eq. \ref{eq:C} results in the minimum-volume solution of the two components ($m_\text{insul}/\rho_\text{insul}+m_\text{sink}/\rho_\text{sink}$), rather than minimum-mass. 

\begin{figure}[t]
	\centering
	\includegraphics[width=\figwidth]{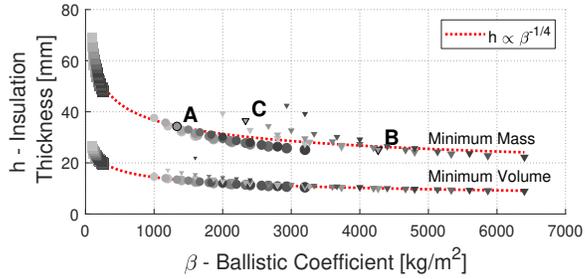}
	\caption{Insulation Thickness Correlation \label{fig:thermalCorrelate}}
\end{figure}

Figure \ref{fig:thermalCorrelate} illustrates the optimal insulation thickness for each simulated probe, both minimum mass and minimum volume optimums. As the descent time, and therefore the heat load, depends primarily on the ballistic coefficient, we again see a data collapse with respect to $\beta$. For thin shells, Eq. \ref{eq:optim} reduces to an insulation thickness proportional to $\beta^{-1/4}$: 

\begin{equation}
\label{eq:b14}
h_\text{insul,optim} \approx \sqrt{K} \propto \sqrt{t_\text{descent}} \propto \beta^{-1/4}
\end{equation}

Most tested probes fall along this $\beta^{-1/4}$ relation, with the exception of the smallest diameters that challenge the thin shell limit. All our tabulated probes (A, B, etc.) use the minimum-mass solution. 

\subsection{Payload Mass Fraction \label{sub:payload}}
In our methodology, the payload mass is simply that left over after subtracting the subsystem masses:
\begin{equation}
\label{eq:massPayload}
m_\text{payload} =  m - m_\text{insul} - m_\text{sink} - m_\text{vessel}
\end{equation}

Combining Eq. \ref{eq:massInsulAndSink} and \ref{eq:optim}, we can first derive the mass fraction dedicated to the thermal system. Using thin-shell assumption, $\dot{Q}_\text{internal} \ll \dot{Q}_\text{external}$, and a constant dominating drag coefficient for the thermally significant subsonic descent portion, we reach the following scaling for our simulation inputs and material choices:
\begin{align}
\frac{m_\text{insul} + m_\text{sink}}{m} &\propto  \frac{1}{\beta_\text{\phantom{B}}^{5/4}C_D} \nonumber \times \left(1-\frac{h_\text{vessel}}{r} \right)^2 \sqrt{f_\text{sink}\eta_\text{insul}\frac{\rho_\text{insul} k_\text{insul}}{\Delta h_\text{sink}}} \nonumber \\
&\propto \frac{C_D^{1/4}}{m_\text{\phantom{B}}^{5/12} \rho_\text{bulk}^{5/6}} \propto \frac{C_D^{1/4}}{d_\text{\phantom{B}}^{5/4} \rho_\text{bulk}^{5/4}} 
\label{eq:massFracThermal}
\end{align}
where again ${h_\text{vessel}}/{r} = \sqrt{[f_\text{vessel} P_\text{atm}(0)]/[0.365 \eta_\text{T} E_\text{vessel}]}$. Of note, while the insulation thickness is dependent only on the ballistic coefficient and material properties (Eq. \ref{eq:b14}), the total mass of the thermal system depends on a coupling of multiple inputs.

Similarly, inverting the pressure vessel buckling criterion Eq. \ref{eq:pressureVessel} produces a mass fraction correlation for the pressure vessel, again assuming thin-shells:
\begin{equation}
\label{eq:massFracPV}
\frac{m_\text{vessel}}{m} = \frac{3\rho_\text{vessel}}{\rho_\text{bulk}}\frac{h_\text{vessel}}{r} \propto \frac{1}{\rho_\text{bulk}}
\end{equation}

The payload mass fraction of the probe (e.g. the efficiency of the design) is therefore maximized for probes that are:
\begin{enumerate}
	\item \emph{Large} - Larger diameters $d$ dilutes the system mass of thermal system ($d^{-5/4}$ in Eq. \ref{eq:massFracThermal}), as the surface area to volume ratio decreases. Small probes, by necessity, cannot take advantage of this effect.
	\item \emph{Streamlined} - Smaller drag coefficient $C_D$ lowers the thermal load for any given sized probe ($C_D^{1/4}$ in Eq. \ref{eq:massFracThermal}).
	\item \emph{Dense} - Larger bulk density $\rho_\text{bulk}$ dilutes the mass of both the pressure vessel and thermal system ($\rho_\text{bulk}^{-5/4}$ and $\rho_\text{bulk}^{-1}$ in Eq. \ref{eq:massFracThermal}-\ref{eq:massFracPV}).
\end{enumerate}

\section{Probe Design} 
\subsection{Mass Tradeoffs}
All three of the maximum payload paradigms discussed above (Sec. \ref{sec:results}.\ref{sub:payload}) also act to shorten the probe descent time, thereby creating a strong tradeoff in the system design of Venus atmospheric probes. Ideally, a Venus probe would have both a large payload \emph{and} a long descent time for science collection and transmission - however this necessarily increases the total mass of payload, structure, and thermal protection. The probe trajectory, therefore, must be carefully selected for the given science constraints of the mission. Small payloads may still be delivered at small mission mass.

\begin{figure}[t]
	\centering
	\includegraphics[width=\figwidth]{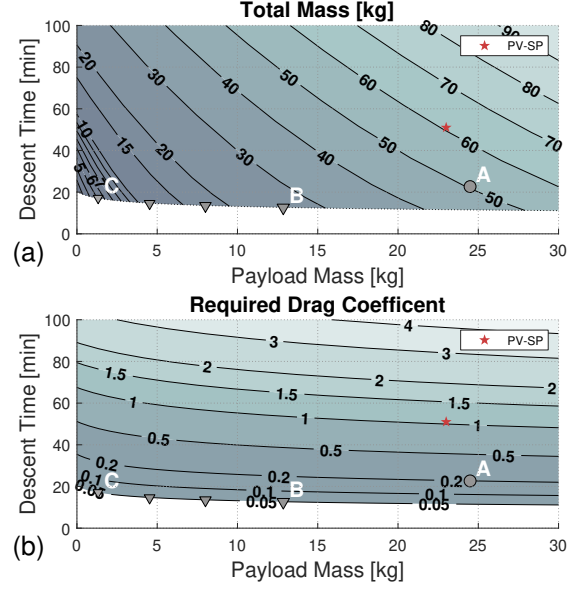}
	\caption{Mass Costs for a target Payload and Descent Time ($\rho_\text{payload}$ = 0.7 g/cm\textsuperscript{3}) \label{fig:payload}}
\end{figure}

Figure \ref{fig:payload} provides summary data to assist the probe designer with this tradeoff. For a Venus probe of specified mission definition (e.g. payload mass and descent time),  we derive the total mass cost and required drag coefficient. All parameters in this figure are solved iteratively using the expressions derived in Sec. \ref{sec:results}\ref{sub:payload} for a payload density of $\rho_\text{payload}=0.7$ g/cm\textsuperscript{3}. 

Figure 10 includes a data point for the PV small probe based on published mass and descent time data. Unfortunately, the published PV data does not include a direct mass estimate that can be compared to our prediction of $\approx23$ kg payload mass as shown. However, \citet{bienstock2004pioneer} does claim a value of 5 kg of science instrument mass, which is a reasonable ~20\% mass fraction of our payload mass that must also include structure, power, avionics and communications hardware. 

The total mass cost in Figure \ref{fig:payload} can be subdivided into two components, a (1) initial cost for descending to Venus without any spare mass allocated for payload, and a (2) marginal cost for either increasing the payload or the descent time:

\subsubsection{Initial Mass Cost at Zero Payload}
Following the y-axis in Fig. \ref{fig:payload}a, the zero-payload cost increases with larger drag coefficients. Combining Eq. \ref{eq:massPayload}-\ref{eq:massFracPV}, this scaling for a constant $C_D$ can be solved as:
\begin{gather}
m \big|_{m_\text{payload}=0} \propto  C_D^{3/5} \left(1-\frac{3\rho_\text{vessel}}{\rho_\text{bulk}\big|_{m_\text{payload}=0}}\frac{h_\text{vessel}}{r}\right)^{-12/5} \times \nonumber \\ 
\left(\rho_\text{bulk}\big|_{m_\text{payload}=0}\right)^{-2} \left(f_\text{sink}\eta_\text{insul}\frac{\rho_\text{insul} k_\text{insul}}{\Delta h_\text{sink}}\right)^{6/5} \left(1-\frac{h_\text{vessel}}{r} \right)^{24/5}
\label{eq:zeroPayloadMass}
\end{gather}
where $\rho_\text{bulk}\big|_{m_\text{payload}=0}$ is the bulk density at zero payload, dependent only on material properties. 

Density $\rho_\text{bulk}\big|_{m_\text{payload}=0}$ can be derived by noting in the zero-payload case, the thermal system volume is the difference between the total and vessel volumes:
\begin{gather}
\rho_\text{bulk}\big|_{m_\text{payload}=0} = \frac{1}{\mathcal{V}} \left[\mathcal{V}_\text{vessel}\rho_\text{vessel}+(\mathcal{V}-\mathcal{V}_\text{vessel})\rho_\text{thermal}\right] \nonumber\\
= \left(1-3\frac{h_\text{vessel}}{r}\right) \frac{2\rho_\text{sink}\rho_\text{insul}}{\rho_\text{sink}+\rho_\text{insul}}+3\rho_\text{vessel}\frac{h_\text{vessel}}{r}
\end{gather}
At the drag coefficients tested, the zero-payload cutoff occurs at 2.1 kg ($C_D=0.05$), 4.9 kg ($C_D=0.2$), and 29 kg ($C_D=4$). Such masses represent an ultimate minimum for vanishingly small (i.e. grams of circuit board) payloads given our material choice and model.

\subsubsection{Marginal Costs of Payload and Descent Time}
For each kilogram of increased payload, the mass of the probe increases by more than a kilogram as the probe mass cascades into the system mass. Similarly, increasing the descent time from any design point also increases the total mass. To derive a rule of thumb, we can fit a linear relation to the roughly parallel contour lines in Fig. \ref{fig:payload}:

\begin{equation}
m \approx k_1 (t_\text{descent}-k_0) + k_2 m_\text{payload} \label{eq:linearcost}
\end{equation}
where costs $k_1\approx 0.30$ kg/min, $k_2 \approx 2.4$ kg/kg, and $k_0 \approx 15$ min for the range of payload masses less than 10 kg and our chosen material properties. For any given payload size, the lowest plausible drag coefficient ($C_D=0.05$) creates a cutoff limit for shrinking the total mass and descent time. 

\subsection{Point Designs}
Carrying on to specific examples, the point designs A, B, and C illustrate the compromises required for shrinking the total mass: both the descent time and payload mass must necessarily drop. Again, Table \ref{tab:probes} enumerates their parameters in detail.

Probe A at diameter $d=50$ cm approximates the Pioneer Venus Small Probe, with deceleration module removed to hasten its descent to the surface to only 23 minutes. Weighing 52 kg, this probe can carry a little under half its mass as payload given the reduced thermal load ($m_\text{payload}/m = 0.47$). It's payload is slightly larger than the estimate for PV-SP, but delivers it with less total mass.

Probe B at $d=40$ cm weighs half of Probe A, so one would expect much less payload could be carried as the effective thermal load increases with larger surface area to volume. However, by adding the cowling to reduce the drag, we maintain a similar mass efficiency. The ratio of total mass to payload mass is maintained ($m_\text{payload}/m = 0.48$).

Probe C represents an extreme design of $d=25$ cm, where the streamlined 5.7 kg probe only carries 1.3 kg of payload ($m_\text{payload}/m = 0.23$). The mass efficiency is necessarily compromised to keep the total mass low, and continued reduction of the total mass will drop the payload to zero. 

Intriguingly, the zero-payload mass for a streamlined $C_D=0.05$ probe is within 3U cubesat range ($m<4$ kg). While a Venus descent probe by no means needs to fit within cubesat design constraints, it is indicative that probes of this size are perhaps plausible for Venus. Such an atmospheric probe would require a kilogram-scale payload, and science collection and transmission performed in approximately 20 minutes.

\subsection{Material Selection \label{sub:material}}
For each of the materials in Table \ref{tab:materials}, improved properties would naturally further miniaturize the design points. The relevant selection criteria can be determined by inspecting our derived scaling laws:
\begin{itemize}
	\item \emph{Pressure Vessel Mass}: Scales as $\rho_\text{vessel}/\sqrt{E_\text{vessel}}$ (Eq. \ref{eq:massFracPV}), so lightweight stiff materials are preferred.
	\item \emph{Thermal System Mass}: Scales as $\sqrt{\rho_\text{insul} k_\text{insul}/\Delta h_\text{sink}\eta_\text{insul}}$ (Eq. \ref{eq:massFracThermal}), so lightweight insulators, low leak rates, and high heat of fusion heat sinks are preferred. As the insulation resides inside the pressure vessel in our model, a thicker vessel (lower stiffness $E_\text{vessel}$) also slightly decreases the thermal load by reducing the insulation area, though at the cost of higher payload density.
	\item \emph{Zero-Payload Initial Mass}: Scales with a complex interplay of parameters, where denser materials generally lower the mass as $(\rho_\text{bulk}\big|_{m_\text{payload}=0})^{-2}$ to leading order and better thermal materials as $(\rho_\text{insul} k_\text{insul}/\Delta h_\text{sink}\eta_\text{insul})^{6/5}$ (Eq. \ref{eq:zeroPayloadMass}). 
\end{itemize}
Accordingly, given otherwise equivalent materials of the same functional expression ($\rho/\sqrt{E}$ or $\rho k$, etc.), a denser material will lower the zero-payload initial mass cost without increasing marginal costs. Moreover, insulation effectiveness is an especially strong driver: a hypothetical insulator of half the conductivity would reduce the mass of a zero-payload probe by $56\%$, as well as the mass fraction of the thermal system by $29\%$. Similar sensitivity will be realized from improvements or reductions in the assumed parasitic heat flows due to electrical feedthroughs in the insulation.

This mass sensitivity to the material choice also acts as a proxy for other miniaturized designs of different system architectures. \citet{lorenz1998design} reports probes on the order of $1$ kg by removing the pressure vessel and relying on the thermal capacity of the probe itself. The constraints on such a payload eliminate much of the protection mass included in our model, driving $h_\text{vessel}/r$ to zero and increasing the effective sink performance $\Delta h_\text{sink}$. 

In summary, further mass savings can be obtained through a variety of factors: improving the thermal materials, hardening the payload, or streamlining the probe to accelerate the descent. Depending on the mission requirements, different methods should be employed to minimize the mass. 

\section{Conclusion}
Venus drop probes are more easily accommodated as secondary mission payloads, either as single probes or in bulk, if the internal components and protection mass can be miniaturized. In this study, we have found that probes of substantially lower mass than Pioneer Venus appear plausible from a thermo-mechanical perspective, provided that streamlining is used to shorten the descent time through the atmosphere. Falling faster mitigates the timeline that heat conduction can occur, thereby reducing the heat load and associated thermal system mass. 

However, fulfilling any science objective with a Venus probe requires adequate time to both collect and transmit data. Accordingly, the system design of a probe must balance the descent timeline with the payload size. For example, one of our proposed 5.7 kg probe designs carries 1.3 kg of payload at an accelerated 17 minute descent. Our kilogram-scale cutoff is already consistent with point designs from the literature \cite{huntress2011soviet,bienstock2004pioneer,lorenz1998design,wells2004atmospheric}, most similar in size to the VEVA imaging probes \cite{kerzhanovich2003low}. Future work is need to match our mass tradeoff analysis with specific payloads, science goals, and transmission rates. More complex cases are also possible: for example, a variable drag or lift device that slows the descent at specific altitudes (such a guided aerosonde \cite{matthies2014venus}) would incur further thermal penalties, especially in the final 5 km near the surface, but lower the data rate requirement. Alternatively, a probe that is only intended to collect data in the lower clouds between from 50-45 km faces an easier thermal environment and miniaturization opportunity. 

Our model uses simplified conservative approximations to ensure robustness of the results, so there is potential for higher fidelity engineering designs to yield further reductions in probe mass. Specifically, we handicap the insulation effectiveness, rely only on the heat capacity of the phase-change heat sink, and make no attempt to harden any payload for the thermal environment. For our conservative model, we derive analytic expressions for optimizing the mass of insulation and heat sink for any given design point. For vanishingly small payloads, the total mass of the probe is still in the low kilograms simply due to the protection mass required. Such designs represent the ultimate limit for Venus probes of conventional payloads and existing materials. 

According to the scaling laws derived in our study, research efforts should prioritize the following areas obtain future improvements in probe designs:
\begin{itemize}
	\item \emph{Streamlined cowlings:} High ballistic coefficients allow small probes of little heat capacity to descend quickly, mitigating the limiting conductive thermal load. To leading order, the mass fraction of the thermal system grows with $C_D^{1/4}$ (Eq. \ref{eq:massFracThermal}).
	\item \emph{Highly integrated thermal structural designs:} Compact packing allows denser probes, which further quickens the descent time for a given probe size. To leading order, the mass fraction of the thermal and pressure systems grow with $\rho_\text{bulk}^{-5/4}$ and $\rho_\text{bulk}^{-1}$ respectively (Eq. \ref{eq:massFracThermal}-\ref{eq:massFracPV}).
	\item \emph{Small, low-mass payloads:} Each increased kilogram of payload cascades into more than a kilogram of added thermal and pressure compensation. For our selected parameter set, payload mass is inflated by 2.4 times into the total mass (Eq. \ref{eq:linearcost}).
	\item \emph{Data transmission and compression:} Advanced data communication allows more science to be obtained for an expected short descent time. Roughly, every three minutes of descent requires an additional kilogram of mass (Eq. \ref{eq:linearcost}).
	\item \emph{Improved material properties:} High temperature insulation and low-leak passthroughs are an especially strong design driver. Halving the specific conductivity of the insulation more than halves the probe mass (Sec. \ref{sec:results}:\ref{sub:material}).
	\item \emph{High temperature and pressure electronics:} The ability to rely on the heat capacity and pressure rating of internal components replaces unneeded protection mass with components that directly enhance science return (see \citet{lorenz1998design}).
\end{itemize}

Finally, although our analysis assumes a probe dropped from a jettisoned heat shield at 65 km altitude, we expect similar minimum-mass limits to apply for a sonde dropped from a Venus aerial platform flying at or above 50 km. The thermal environment only begins becoming challenging below this altitude, and the sonde reaches terminal velocity so quickly that a slower initial aerial velocity would not appreciably affect the analysis.

\section*{Funding Sources}
The research described in this paper was funded by the Jet Propulsion Laboratory, California Institute of Technology, under contract NNN12AA01C with the National Aeronautics and Space Administration.

\section*{Acknowledgments}
The authors wish to acknowledge the Venus Bridge and Venus Aerial Vehicles study groups for motivating this research and providing thoughtful feedback throughout its development. Specific thanks to James Cutts for providing a leadership role in both study groups, facilitating discussions with the larger Venus community, and lending his system engineering expertise.

\bibliographystyle{new-aiaa}
\bibliography{biblioVenus}

\end{document}